\documentclass{article}
\usepackage{graphicx} 
\usepackage{url}
\usepackage{authblk}
\title{Multilayer networks describing interactions in urban systems: a digital twin of five cities in Spain}
\author[1,2]{Jorge P. Rodríguez}
\author[3,4]{Alberto Aleta}
\author[3,4,5]{Yamir Moreno}
\affil[1]{Instituto de Física Interdisciplinar y Sistemas Complejos (IFISC), CSIC-UIB, Palma de Mallorca, Spain}
\affil[2]{CA UNED Illes Balears, Palma, Spain}
\affil[3]{Institute for Biocomputation and Physics of Complex Systems (BIFI), University of Zaragoza, Zaragoza, Spain}
\affil[4]{Department of Theoretical Physics, Faculty of Sciences, University of Zaragoza, Zaragoza, Spain}
\affil[5]{CENTAI Institute, Turin, Italy}

\begin{document}

\maketitle
\begin{abstract}
Networks specifying who interacts with whom are crucial for mathematical models of epidemic spreading. In the context of emerging diseases, these networks have the potential to encode multiple interaction contexts where non-pharmaceutical interventions can be introduced, allowing for proper comparisons among different intervention strategies in a plethora of contexts. Consequently, a multilayer network describing interactions in a population and detailing their contexts in different layers constitutes an appropriate tool for such descriptions. These approaches however become challenging in large-scale systems such as cities, particularly in a framework where data protection policies are enhanced. In this work, we present a methodology to build such multilayer networks and make those corresponding to five Spanish cities available. Our work uses approaches informed by multiple available datasets to create realistic digital twins of the citizens and their interactions and provides a playground to explore different pandemic scenario in realistic settings for better preparedness.
\end{abstract}

\section{Background \& Summary}

Understanding contact patterns is crucial for modeling the spread of infectious diseases within a population. Over the past decade, mathematical and computational models have advanced from the traditional homogeneous approximation, incorporating complexities that extend beyond mere knowledge of who interacts with whom. These models now aim to account for critical details, including when, where, under what circumstances, and the duration of these interactions \cite{Mossong2008Mar}. Such nuances can profoundly influence the outcomes of epidemiological models, yet the accessibility to this information often remains limited.

One common method for representing interactions is through the use of contact matrices, which contain interaction rates among different cohorts, typically categorized by criteria such as age \cite{Manna2023Jun}. These matrices can be constructed using data from surveys \cite{hoang2019systematic}, census data \cite{fumanelli2012inferring}, or a combination of both \cite{prem2017projecting, mistry2021inferring}. However, this approach treats all individuals within a given group as equivalent. To capture individual heterogeneities, researchers resort to data-driven approaches \cite{Iozzi2010Dec, Grefenstette2013Dec, Gallagher2018Jun} or to create contact networks in which each node represents an agent, and the links encode their interactions \cite{Pastor-Satorras2015Aug}.

There are several methodologies for collecting data and building interaction networks. The simplest approach is to carry out surveys, a technique that is adequate for relatively small populations and situations where contacts are easily identifiable, such as sexual interaction networks  \cite{Klovdahl1994Jan, Jolly2001Sep, Liljeros2003Feb}. However, there are some diseases, like those that are airborne, for which the potential number of contacts is much larger and harder to identify.

For these scenarios, high-resolution data collection can be achieved using wearable sensors \cite{Genois2018Dec}, but logistical constraints limit their applicability to specific settings \cite{Smieszek2016Dec} or small areas \cite{Ozella2021Dec}. During the recent COVID-19 pandemic, digital contact-tracing apps were rapidly deployed, offering the potential to build large-scale networks. However, low participation rates, technical issues, and privacy concerns hindered the success of these apps \cite{Pandit2022Jul,Barrat2021May,Burdinski2022Dec,rodriguez2021population}. An alternative approach involves inferring synthetic contact networks from survey data used to create contact matrices, but since most surveys only stratify individuals by age, many socio-economic characteristics of the population remain unaccounted for \cite{Aleta2020Jul}.

In response to these challenges, we propose the use of publicly available, high-resolution data to construct synthetic multilayer networks that faithfully replicate key features of urban areas. In multilayer networks, the system is divided into multiple layers, each representing a specific type of interaction \cite{kivela2014multilayer, aleta2019multilayer}. This framework allows us to naturally account for different settings in which human interactions occur and also their distinct impacts on pathogen transmission. In this work, we present the multilayer networks that describe individual-level interactions in five Spanish cities, identifying six distinct interaction contexts or layers: households, schools, universities, nursing homes, workplaces, and the broader community. In addition to their utility in studying disease spreading across diverse urban settings and measuring the distinct impacts of characteristic features unique to each city, they are also a versatile platform that may be used to investigate other spreading processes extending beyond epidemics.

\section{Methods}
\begin{table}
    \centering
    \begin{tabular}{|c|c|c|c|c|c|}
    \hline
     City & $D$ & $N$ & $N_{NH}$ & $N_{EU}$ & $N_{TOT}$\\ \hline \hline
     Barcelona & 10 & 1,666,530 & 16,456 & 129,644  & 1,812,630\\ \hline
     Valencia & 19 & 801,545 & 75,708 & 2,193 & 879,446 \\ \hline
     Sevilla & 11 & 701,455 & 2,655 & 64,371 & 768,491\\ \hline
     Zaragoza & 12 & 681,877 & 6,905 & 17,396 & 706,178\\ \hline
     Murcia & 8 & 459,403 &  1,335 & 31,800 & 492,538\\ \hline
    \end{tabular}
    \caption{Sizes of the studied cities. $D$ is the number of districts considered by the census, $N$ is the population according to the available data, $N_{NH}$ is the population living in nursing homes, $N_{EU}$ is the number of external university students and $N_{TOT}=N+N_{NH}+N_{EU}$.}
    \label{tab1}
\end{table}

\subsection{Geographical extents}

To limit the geographical scope of our study, we employed the administrative boundaries of each municipality within the respective cities under consideration. Within these municipal limits, we utilized census districts as the primary geographical units assigned to individuals. These census districts were defined by shapefiles available in the cartography of the 2011 census provided by the Spanish ``Instituto Nacional de Estad\'istica'' (INE).

\subsection{Demography}

To construct the synthetic population required for inferring multilayer contact networks, we created a population that accurately mirrored the observed distribution of citizens' age and sex across various districts within each city. We sourced recent demographic information from local census data, primarily obtained from the municipalities' online open datasets as of January 1, 2020. This data was readily available for Barcelona, Valencia, Sevilla, and Murcia. For Zaragoza, we utilized continuous census data provided by the ``Instituto Nacional de Estadística'' (INE).

The diverse origins of these datasets, spanning local, regional, and national administrative units, resulted in varying age groupings, typically organized into 5-year cohorts, with variations in the lowest and highest age categories. Specifically, the most elderly age group was designated as +100 for Barcelona and Zaragoza, +90 for Valencia and Sevilla, and +85 for Murcia. Moreover, Murcia subdivided the '0-4' age category into '0' and '1-4'. We decided not to standardize the age-groups, aiming to preserve as much information as possible for each city. This decision also allows for assessing the significance of such detailed age information. If necessary, it can be easily aggregated to create a standardized model.

Using the census statistics, we created a synthetic population, specifying the age, sex, and residential district of each individual. For layers that required precise ages (such as university and school layers), we allocated ages with a 1-year resolution for individuals under 30 years old using an interpolation method. This method involved considering an age, say 'A', and then calculating the set of ages: {A-2, A-1, A, A+1, A+2}. We used the number of ages in this set within each binned age-cohort as a weighting factor to estimate the expected number of individuals at that specific age. For instance, for an individual aged 29, the contribution from the 25-29 age cohort would be 3/5, while the contribution from the 30-34 cohort would be 2/5.

Once this synthetic population was created, we compared the total population count within each district with the total population by age and sex in the same district. This validation process matched in Valencia, Sevilla, Zaragoza, and Murcia, and thus, these populations were retained. However, in the case of Barcelona, the total population count, when summed across ages and sexes, did not align with the reported total population in the district. We hypothesize that it was caused by the absence of information on age and sex for some individuals. To address this, we randomly sampled from the synthetic population to maintain the total number of individuals reported, resulting in the final population considered for the synthetic city.

\subsection{Households layer}

To create the households layer, we used various statistical parameters within each district, including the number of households, average household size, the distribution of household sizes, and the structure of these households. Additionally, for households including adult couples, we considered the average age difference between cohabiting partners.

Our approach began by calculating a scaling factor, representing the ratio between the expected population (calculated from the product of the average household size and the number of households) and the actual population (as detailed in the Demography section). In most cases, this factor was close to 1, necessitating no further adjustments. 

Household sizes were obtained as integer values for smaller sizes but transitioned to categorical sizes for larger households, defined as 9 or more people in Barcelona, and 6 or more people in Valencia, Sevilla, Zaragoza, and Murcia. For these larger households, we estimated their average size, ensuring that the district's overall average household size remained unchanged. This average value was then assigned to these larger households.

\begin{figure}[hbt]
    \includegraphics[width=1.2\textwidth]{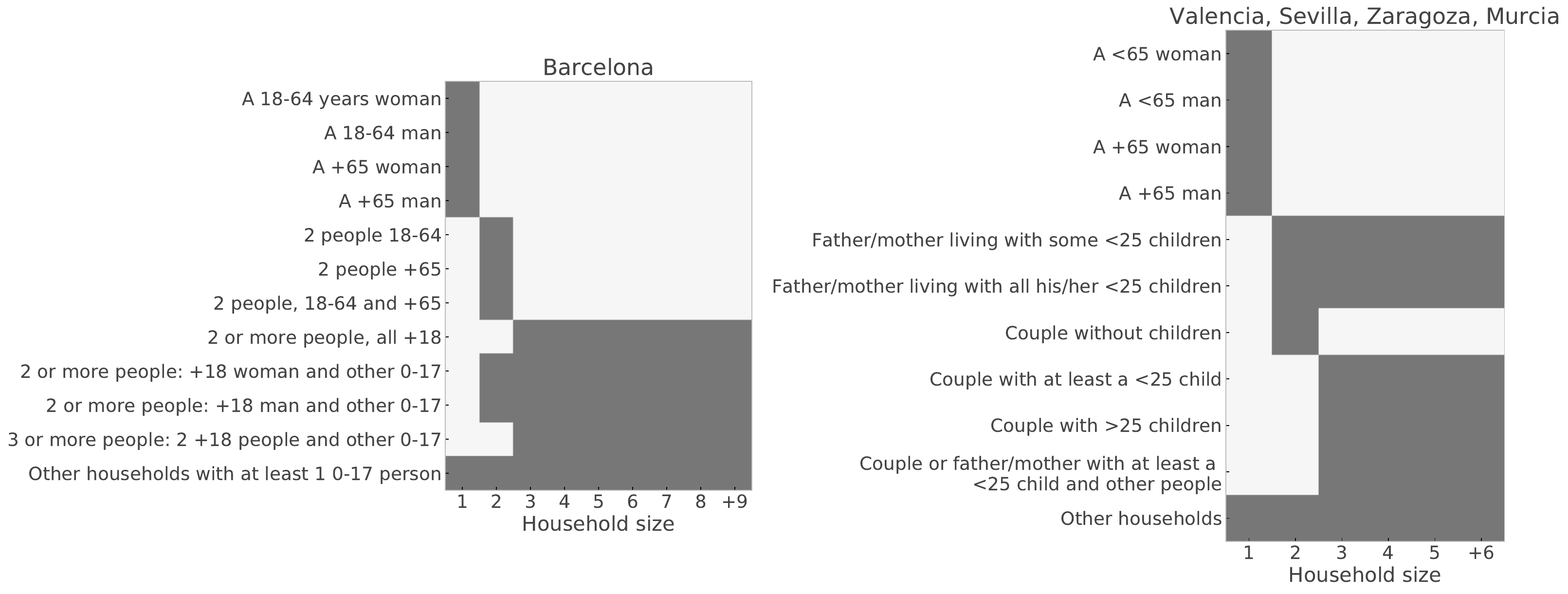}
    \caption{Compatibility matrices between the structure of the household and its size. Darker entries represent the combinations that are allowed.}
    \label{fig1}
\end{figure}

To account for the specific household structures, which encompassed broad age cohorts and gender, we associated each structure with household size. We achieved this by using available categories for each city to create a compatibility matrix (see Fig. \ref{fig1}).

Using this information, our household creation algorithm worked as follows. First, for each census district, the average size for unbounded households (e.g., +9 for Barcelona, +6 for Valencia) was calculated and rounded to the nearest integer. In the unlikely case of these sizes being smaller than the minimum boundary, then these households were assigned that boundary. If the average exceeded 20, households were divided into patches of 20 individuals each. 

Subsequently, the algorithm generated two vectors for each household: one describing its structure and another detailing its size. Given that not all households in the district had available statistics, we generated household sizes based on the distribution of those with available information while keeping the district's population fixed (see Demography). 

For household structures, we allocated individuals according to the profiles present in the data (for Barcelona, sex and age cohorts 0-17, 18-64, +65 and, for Valencia, Sevilla, Murcia, and Zaragoza, sex and age cohorts 0-24, 25-64 and +65). We started with households of size 1; for all the cities, these structures of households corresponded to 4 combinations: woman/man with ages 18-64 or +65. We generated unoccupied households and aimed to fill all of them with synthetic individuals according to their profiles. If, for a given structure, there were no remaining individuals to allocate in the household, we kept that household pending for a secondary iteration of the algorithm performed after assigning the households of size 2. 

We then followed the same approach for the households of size 2, keeping the distribution of structures compatible with their sizes (Fig. \ref{fig1}). Additionally, we considered whether the household nuclei were composed of a couple of woman-man or not (statistics also available in the age-difference dataset). If this was the case, we also considered the distribution of age differences. After assigning the households of sizes 1 and 2 to their structures and individuals, some pending households could not be filled with the synthetic population, so they were assigned to the category 'Other' (Valencia, Sevilla, Zaragoza, and Murcia) or 'Other households with at least a 0-17 person' (Barcelona), which in the latter corresponded to households of size 1 and 2 with individuals with 0-17 age. 

Lastly, we repeated the same procedure for households with sizes larger than 2. After that, if there were still some unoccupied households that could not be filled according to the distribution of the structures, they were completed randomly with the remaining individuals who did not have an assigned household.

\subsection{School layer}

The creation of the school layer necessitated detailed information about educational institutions, including the levels they offered and their respective districts, as well as data on student groups (such as the number of groups per level and their sizes) and the number of students per level or group. This data was sourced from regional education councils. However, the data was not consistently formatted, and variations in resolution existed among the datasets.

For this layer, we focused on four key educational programs: infant (0-5 years), primary (6-11 years), secondary (which in Spain encompasses compulsory education for ages 12-15 and baccalaureate for ages 16-17), and work training (age 16 and above). Notably, work training programs exhibited a wide range of student ages, necessitating special consideration. In contrast, for the other educational programs, we assumed that all students fell within the expected age ranges for their respective levels.

Discrepancies emerged in the number of enrolled children and those included in our synthetic population for specific age groups. This issue did not pose a problem when the synthetic population numbers were higher, given the incorporation of recruitment rates below 1. However, in cases where the official enrollment statistics reported higher numbers, we scaled them to match the figures within the synthetic population. This divergence might arise from variations in data collection times between demographic and educational datasets, as well as the registration of children from other municipalities.

To perform this scaling, we also considered that the minimum legal working age in Spain is 16 years old. Accordingly, we extracted the employment rate for individuals aged 16-19 as of January 2019. This information helped us estimate the number of individuals within this age range who were not in the job market and, therefore, were expected to enroll in educational programs. We then compared these estimates with the numbers reported in the statistics, aiding in the alignment of data from different sources and resolving discrepancies.

Then, our goal was to create synthetic educational centers that included students organized into groups according to different educational levels, each overseen by a teacher. We then established connections between students within the same group and between teachers within the same center, whenever specific center-level information was available. In cases where detailed center-level data was lacking, we connected teachers associated with the same educational program. This general framework guided our methodology, which was adapted and refined based on the availability, format, and resolution of data for each city. In the following, we describe the data and the processing methods employed for each city.

\subsubsection{Barcelona}

For the case of Barcelona, the statistics were available at the census district level for the academic course 2019-2020. There were available datasets associated with the first (0-2) and second (3-5) cycles of infant education, primary education, secondary education, baccalaureate, and medium and advanced work training programs. The file reports the number of students specified, for each level (generally associated with age), and the number of registered students in the centers of each census district, split amongst different management modes (regional educational council, local authorities, or private). 

Additionally, we had data on the centers and groups, which included the number of centers per district with students in each educational program and the number of groups, although not specified by level. Our objective was to create synthetic educational centers with groups assigned to specific levels. To do so, we divided the process in two steps:

\begin{enumerate}
    \item Group creation: We began by looping through different districts and management modes. For each combination, we ensured that the number of groups was at least as large as the number of levels within that educational program. In that case, we created the groups and randomly assigned them a level while making sure that at least every level had a group. If the number of groups was lower than the number of levels, we assumed that groups encompassed different levels, which was more common in the first cycle of the infant program.

    \item Center and group assignment: We then randomly assigned a center to each group, followed by the distribution of the number of students (scaled from statistics) among these groups and their respective centers.
\end{enumerate}

To populate these centers, we matched individuals in our synthetic population based on their ages and districts to the students' ages and census districts. Initially, we filled the groups with individuals of the same age living in the same district as the center. Subsequently, any remaining vacant student positions were filled with individuals from neighboring districts. We employed a gravity law approach for this purpose, with the probability of transitioning between districts being proportional to the squared inverse of the great circle distance between their centroids.

The procedure was similar across different programs, except for the medium and advanced work training programs. For those cases, we utilized an additional dataset that provided the number of individuals registered in each district, by management mode and at each level, with specific ages (measured in years). This information allowed us to create a more accurate representation of the individuals enrolled in these programs.

To complete the educational centers, we assigned a teacher to each group. We selected individuals from the synthetic population aged between 30 and 70 years old who were not already assigned to any school group (note that some individuals older than 30 may be registered in the work training programs). The teacher assignment process mirrored that of students, ensuring that teachers belonged to the same school group and educational center. We followed the same algorithm to assign the teachers to their groups in the rest of the cities. Note that in this approach students only interact with students in the same group, as well as with the teacher, while teachers may interact with any other teacher in the same center.

\subsubsection{Valencia}

For Valencia, our data source provided detailed information for the academic year 2019-2020. Specifically, it outlined the number of groups offered and the number of registered students at each level of the considered programs, for each specific educational center. This high-resolution data greatly facilitated the creation of educational centers and groups within our digital city.

However, since the dataset did not include information about census districts, we turned to another dataset that included geographical coordinates for each educational center. Using this information, we assigned a specific census district to each center based on its geographical location and the boundaries of the districts (as described in the ``Geographical Extents'' section). A similar approach was applied to Sevilla, Zaragoza, and Murcia.

In the case of work training programs, we relied on national statistics related to the ages of registered individuals for basic, intermediate, and advanced programs. This approach was also applied in Sevilla, Zaragoza, and Murcia.

In scenarios where the number of registered students in the schools of a district exceeded the number of synthetic individuals in that district, we adopted a strategy to balance these figures. Specifically, we selected registered individuals from districts with the opposite scenario, where there were more synthetic individuals than registered students. This process followed the gravity-based approach described earlier for Barcelona, ensuring a balanced representation of individuals within our synthetic population.

\subsubsection{Sevilla}

In the case of Sevilla, the available data had a lower level of resolution compared to other cities. Specifically, we had access to datasets detailing the number of registered students for each program, with further breakdowns by management modes (public, private, and private concerted centers). Additionally, another dataset provided information on the number of groups associated with each program. These datasets pertained to the academic year 2017-2018.

To address the lack of information at the district level, our approach involved iterating over the educational programs, taking into account the programs offered by each school. It's important to note that for infant programs, there was no differentiation between the first (0-2 years) and second (3-5 years) cycles. Therefore, we assumed that students in the first cycle attended centers exclusively offering the infant program.

To populate the synthetic city, we assigned individuals to groups located within their respective census districts. When allocating individuals to groups, we followed the gravity-based approach for centers that included all students from the same district. This method helped ensure that individuals were assigned to groups in a manner consistent with their geographical proximity, considering both program and district-specific factors.

\subsubsection{Zaragoza}

For Zaragoza, our data sources provided information at the municipality level, with data corresponding to the academic year 2019-2020. Specifically, we had access to the number of students enrolled in each program, along with specifications regarding the management mode of the educational centers (public, private, and private concerted centers). Additionally, data on the number of groups were available, detailing the number of groups per age group and management mode. The information also included details on how many centers offered each program, further categorized by management mode. Within this center description, we found information regarding the types of centers that we matched with the educational programs under consideration.

Then, our approach closely mirrored that used for Seville. We iterated over educational programs, ensuring that each center offered at least one group for each level when possible. Any remaining groups were assigned randomly. Following the group assignment, we distributed registered students among these groups and matched them with individuals in our synthetic population. Lastly, we applied the gravity-based approach, as previously described, in cases where the number of registered students in a district exceeded the population of the district at a specific age. 

\subsubsection{Murcia}

In the last city, Murcia, we encountered data of a similar nature as in Sevilla and Zaragoza, with information available at the municipality level and data corresponding to the academic year 2016-2017. The available data included details on how many students were registered for each program, categorized by two management modes (public and private). Similarly, the number of groups was also reported at this resolution. Additionally, we had access to data on the number of centers, their geographical coordinates, management mode, and type. The center-type information was particularly useful for associating the different educational programs with each center.

To create groups, we iterated over educational programs, ensuring that each center offered at least one group per age when possible. Following this group assignment, we distributed registered students among these groups and matched them with individuals in our synthetic population. As in previous cases, when the number of registered students of a specific age in a district exceeded the number of individuals of that age in our synthetic population, we applied the gravity-based method to balance these figures.

\subsection{University layer}

This layer connects students within the same academic program at both the Degree and Master's levels. We utilized data from the Ministry of Science and Innovation in Spain, which provided information on the number of registered students and the percentage of women in each program at various universities. To ensure geographical relevance, we focused on universities located in the municipality of the city or its metropolitan area. The only exception was the Universidad Politécnica de Cartagena, located in the region of Murcia but outside the capital's metropolitan area.

Next, we incorporated information on the age distribution of students. This data, also obtained from the Ministry of Science and Innovation, categorized students into age and sex groups for the academic year 2019-2020. For Degree students, the age groups were [18-21, 22-25, 26-30, +31], while for Master's students, they were [$<$25, 23-50, 31-40, +41]. Using this data, we constructed vectors indicating the age group of each female and male student in both Degree and Master's programs.

Not all registered students were likely represented in our synthetic population as university students often have high geographical mobility. We estimated the number of registered students present in our synthetic population using two sets of data. First, we used publicly available statistics describing the province of origin of the students registered in each university. This allowed us to estimate the fraction $F$ of students who reside in the same province as their university's location. Then, we obtained data from a private request to the Ministry of Science and Innovation which provided the number of registered university students per municipality. We calculated the ratio $R$ between the number of registered students from the city and the total number of students from any municipality in the province, considering 5 students for municipalities that specified '5 or less'. Consequently, the fraction of registered university students present in our synthetic population was estimated to be $F\times R$. The remaining students were treated as external university students who interacted exclusively within the university layer of our multilayer network.

With all this data in hand, we created synthetic university groups as fully-connected patches. This involved considering the number of registered students in each Degree or Master's program, their gender distribution, and their age distribution based on regional statistics. For programs with more than 50 registered students, we randomly divided them into multiple groups, each consisting of 50 students, plus any remainder. Within these groups, a fraction ($FR$) represented our synthetic population, and we selected individuals with matching ages and genders to populate these synthetic university groups. We also stored the features (age and gender) of the external university students and assigned them to their respective university groups.

\subsection{Nursing homes layer}

In the context of studying infectious disease dynamics, and particularly in light of the recent COVID-19 pandemic, nursing homes are recognized as high-risk locations that warrant inclusion in our synthetic cities \cite{ahrenfeldt2021sex, Koleva2021Dec}. However, determining the registration status of residents in the local census proved to be challenging. To address this uncertainty, we decided to treat nursing home residents as an external population separate from our synthetic population. These residents were considered to interact exclusively within the nursing home layer.

We gathered information about nursing homes and their capacities from the program ``Envejecimiento en red'' (Aging in a network) from the Spanish National Research Council. In particular, for each province, there was available a file detailing the number of nursing homes, their location, and their respective capacities. While the general dataset was updated in 2020, some nursing homes had older updates.

To build this layer, we assumed that the number of residents in each nursing home was equal to its capacity. Additionally, we introduced a caretaker for every 4 residents selected from the synthetic population among those within the legal age limits to work, and who did not have any connections in the school layer (neither as teachers nor students). As previously mentioned, residents were considered as part of the external population, so their age and sex were randomly assigned based on national statistics of residents in nursing homes, obtained from the Instituto Nacional de Estad\'istica (INE). Lastly, each nursing home was assumed to be a fully connected cluster so that all residents and caretakers belonging to the same nursing home could interact with each other.

\subsection{Work layer}

This layer represents interactions within the working environment. To realistically model this layer, we obtained data on both employee attributes and company sizes.

For the employee information, we considered the number of individuals affiliated with Social Security in each city $i$ as of January 2019, $A_i$. It is important to note that this number includes various categories, such as unemployed individuals actively seeking employment. To refine this data, we obtained the occupancy rate $O_i$ which represents the fraction of eligible individuals who were currently employed. Additionally, we calculated the total number $T_i$ of individuals in each city who were eligible to work, meaning they were older than 16 years. Consequently, the number of workers is $O_iT_i$

With all this information, we estimated the number of workers in each city as follows. First, we obtained the number of freelance workers $F_i$ from the Social Security dataset. Second, we defined an activity ratio $r_{W,i}$ as $O_i T_i/A_i$, representing the ratio between the expected number of workers ($O_iT_i$) and the number of individuals registered in the Social Security system ($A_i$). Finally, we rescaled the difference between $A_i$ and $F_i$ with the activity ratio to obtain the final number of active workers in companies in the city $i$, $W_i = (A_i - F_i) r_{W,i}$.

Before proceeding to assign ages and sex to these workers, we excluded those who had already been allocated to other layers, such as teachers in the school layer and caretakers in the nursing homes layer. For the remaining workers, we assigned ages and genders using national statistics from INE for January 2019. These statistics categorized individuals into various age groups: [16-19,20-24,25-29,30-34,35-39,40-44,45-49,50-54,55-59,60-64,+65] (with +65 representing ages 65-69). We randomly selected the sex first, based on the estimated distribution, and then randomly assigned an age group to each worker.

Now that we had our ``virtual'' workers (not yet matched with the synthetic population), we allocated them to companies. Data on company size distributions were available from INE at the province level. This data specified the number of companies within one of the following size ranges: [1-2,3-5,6-9,10-19,20-49,50-99,100-199,200-499,500-999,1000-4999,+5000]. We represented these distributions by assigning the central value to each range, resulting in values [1.5,4,7.5,14.5,29.5,69.5,149.5,349.5,749.5,2999.5]. Notably these distributions followed the same shape across the 50 Spanish provinces, which was described by a power-law distribution $pdf(S) \sim S^{-2}$. Thus, we used this information to generate the companies' sizes using the Python package \emph{powerlaw} \cite{Alstott2014Jan}. The minimum company size was set to 1, and the maximum size was determined by the number of workers who had not yet been allocated.

After generating the size of each company, we randomly assigned workers from the virtual set. For companies with more than 20 workers, employees were distributed randomly among a number of patches equal to $P = 1+S$ (mod 20) patches. Finally, we assigned the company patches to individuals in the synthetic population, generating a fully connected pattern within each patch.

\subsection{Community layer}

Due to limited data availability for this layer, we adopted a data-driven approach based on the contact matrices reported in \cite{prem2017projecting} for the category ``Other locations'' in Spain. These matrices specified contact patterns between individuals grouped into 5-year age brackets up to the age of 80. For older age groups, we extrapolated from the last available age group.

Our modeling approach assumed that community interactions occurred among individuals residing in the same census district. Then, we estimated the total number of links from a specific age group $i$ to another age group $j$, $E_{ij}$ as
\begin{equation}
    E_{ij} = P_i \cdot \frac{C_{ij}+C_{ji}}{2}
\end{equation}

\noindent
Where $P_i$ represents the population in age group $i$, and $C_{ij}$ is the number of contacts between individuals in age group $i$ and those in age group $j$ \cite{Arregui2018Dec}. Then, we simply extracted $E_{ij}$ random links from all the possible links between individuals in age groups $i$ and $j$ and added them to this layer.

\section{Data sources}
The generative process leading to our multilayer networks was informed by datasets which in most cases were publicly available. This section details which datasets were particularly used, together with the details of the final products, \emph{i.e.} the multilayer networks.
\subsection{Scientific literature}
We downloaded the contact matrices for 152 countries \cite{prem2017projecting,cmatrices152}, using the data for Spain and the contacts categorized as 'other locations', to inform the generation of the contacts in the community layer.
\subsection{National level}
We obtained from INE \cite{inecensus2011} the following information at the district level: population by age and sex, home sizes, and home structure.

The Ministry of Science, Innovation and Universities (Spain) provided the number of registered students per university and specific degree \cite{uniscience}. This contributed to the generation of the university layer, together with information on the students' ages per region, obtained from the Ministry of Education, Culture and Sport \cite{unistatsdegree,unistatsmaster}. We also included another dataset after a request to the Ministry of Universities, detailing the number of registered students in BsC and MsC programs per municipality in the academic course 2018-2019.

In the school layer, we considered training program students' ages, obtained from the Ministry of Education, Culture, and Sport \cite{fpstats}.

To create the work layer, we used the number of companies per region and their number of employees \cite{inecompanies}, the results of the national employment survey \cite{ineepa}, both available from INE, together with the statistics of affiliated workers to the Spanish Social Security system by municipality \cite{segsoc}. 

Finally, the nursing homes layer was created using data describing the age and gender of residents in nursing homes \cite{inenursing} from INE, as well as the number of places in each of the nursing homes located at each municipality, from the `Envejecimiento en Red' program \cite{envenred}.

\subsection{Barcelona}

To create the synthetic population, we used the local census detailing how many individuals, by age and sex, were registered in each district from Barcelona City Council \cite{bcncensus}. 

In the school layer, we used a dataset from the Catalan Government describing the number of registered students, units and centers per district \cite{bcnschool}.

\subsection{Valencia}

We obtained from Valencia City Council the population statistics at the district level, specifying ages and sexes \cite{vlccensus}.

We used data on the number of students per level of each program in school, by centers \cite{vlcschool}, the same information for job training programs \cite{vlctrain}, and the list of scholar centers, together with their geographical coordinates \cite{vlccentre}, available from Valencia Regional Government.
\subsection{Sevilla}

We considered the population by age and sex, specified at district resolution, available from Sevilla City Council \cite{sevcensus}.

To create the school layer, we downloaded data describing the number of students, number of centers, and number of units, per level and municipality \cite{sevschool}, and the list of scholar centers, including their coordinates \cite{sevcentre}, provided by Andalucía Regional Government.

\subsection{Zaragoza}

We considered, at the level of municipality, the number of students, number of units, and number of centers per level \cite{zgzschool}, together with the list of scholar centers that included their coordinates, both available from the Arag\'on Regional Government.

\subsection{Murcia}
Considering that districts were divided into sections, we used the population by age, at section resolution for men \cite{murcensusm} and women \cite{murcensusf}.

To create the school layer, we used a dataset detailing at each level and within each municipality, the number of registered students, the number of units, and the number of centers \cite{murschool}, together with the list of centers that included their coordinates \cite{murcentre}.

\subsection{Multilayer networks and metadata}

We are releasing, together with this article, the multilayer networks of each layer in each city, as a link list in a file in CSV format, where each row represents a link and reports the two individuals that it connects.

Additionally, we include three metadata files in CSV format describing the individuals' features: first for the generic population, second for the nursing home residents, and third for the external university students. Each file details the age and sex (male: 0, female: 1) of each individual, together with the census district for the generic population.

\section{Validation}

We conducted an analysis of the macroscopic features of the multilayer network that describes the contact within our digital cities to validate our methodology. First, we computed the average degree for individuals who had interactions on each of the layers, as shown in Table \ref{tab2}. Note that the average number of connections in each layer may vary significantly, and thus interactions should be properly weighted depending on the process being studied \cite{mistry2021inferring}.

\begin{figure}
\centering
\includegraphics[width=1\textwidth]{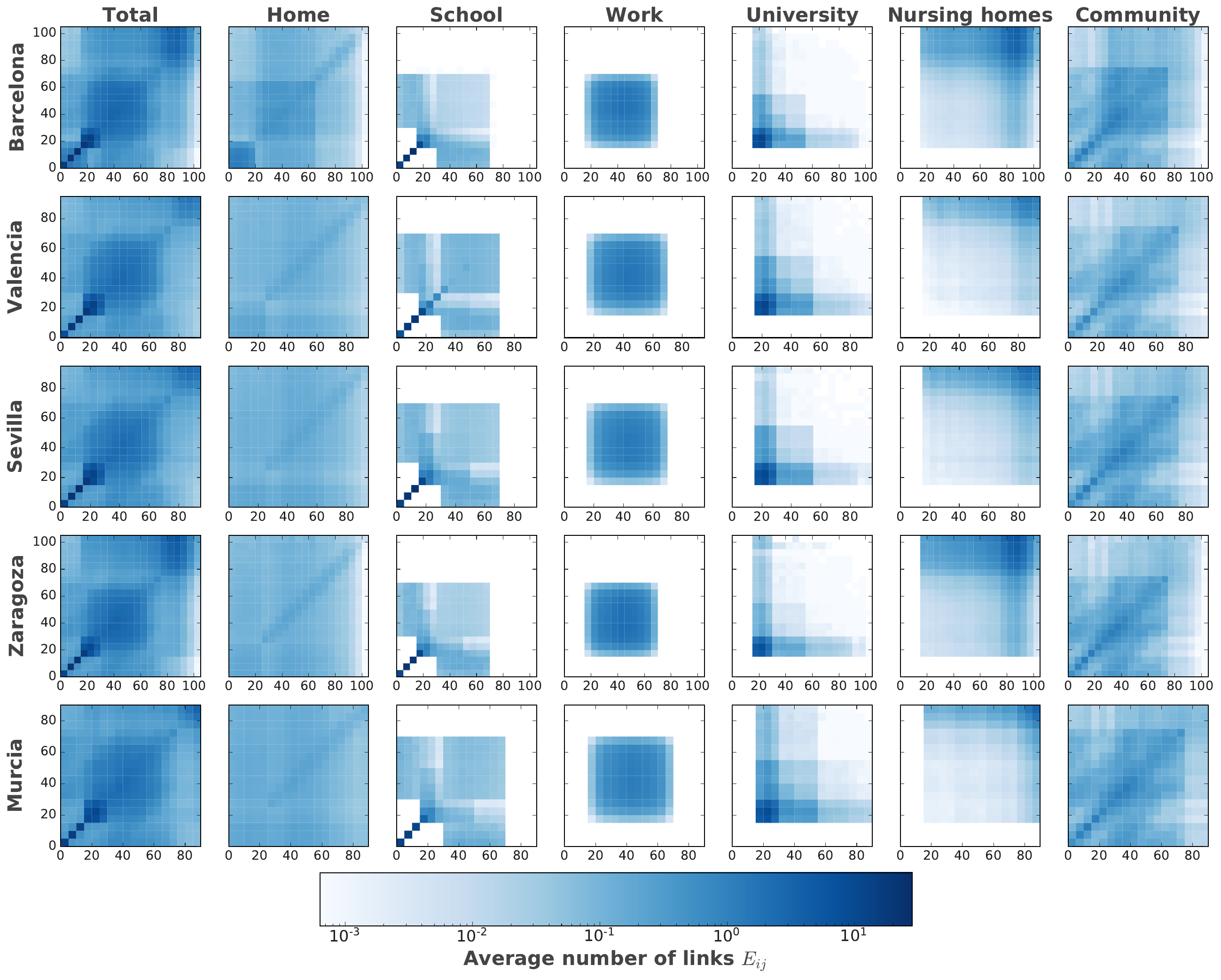}
\caption{Contact matrices in each digital city aggregated (Total) and per layer. Each entry $E_{ij}$ indicates the average number of connections that the individuals of the age represented by row $i$ will have with the individuals in the age group of column $j$.}
\label{fig2}
\end{figure}

Second, we analyzed the contact matrices at an aggregated level and for each layer extracted directly from the multilayer network. The results of this analysis are presented in Fig. \ref{fig2} and are consistent with previous studies analyzing contact matrices \cite{prem2017projecting}.

\begin{table}
    \centering
    \begin{tabular}{|c|c|c|c|c|c|c|c|}
    \hline
     City & $\langle k_H \rangle$ & $\langle k_S \rangle$ & $\langle k_U \rangle$ & $\langle k_{NH} \rangle$ & $\langle k_W\rangle$ & $\langle k_C \rangle$ \\ \hline \hline
     Barcelona  & 3.74 & 26.25 & 45.43 & 139.88 & 14.21 & 4.31\\ \hline
     Valencia  & 2.05 & 24.62 & 45.77 & 143.73 & 15.32 & 3.64\\ \hline
     Sevilla  & 2.26 & 25.87 & 46.36 & 158.32 & 14.33 & 3.72\\ \hline
     Zaragoza  & 2.03 & 23.19 & 46.56 & 149.22 & 18.35 & 3.64\\ \hline
     Murcia  & 2.48 & 22.61 & 46.29 & 145.19 & 12.40 & 3.87\\ \hline
    \end{tabular}
    \caption{Average number of contacts in the synthetic cities per layer: households, $\langle k_H \rangle$; schools, $\langle k_S \rangle$; universities, $\langle k_U \rangle$; nursing homes,$\langle k_{NH}\rangle$; work, $\langle k_W \rangle$; and community, $\langle k_C\rangle$.}
    \label{tab2}
\end{table}

Lastly, these multilayer networks were used to study the spreading of COVID-19 during the first and second waves in Spain. The calibrated models on these networks showed a good agreement with the data extracted from surveillance reports and offered insights on the mechanisms behind the spreading through the population \cite{rodriguez2023digital}.

\section{Discussion}

Synthetic datasets describing social contacts in cities represent valuable tools to inform mathematical spreading models describing for example epidemics \cite{rodriguez2023digital} or opinion evolution \cite{fernandez2014voter}. By following the methodology outlined in this article, we have created synthetic cities across five different autonomous regions in Spain, even when faced with the challenges of varying data availability and formats.

While we acknowledge that the inferred contacts presented in these networks are only a static picture that will change over time, we emphasize the importance of releasing these datasets and providing the code for generating synthetic cities. This not only facilitates updates but also encourages further research and modeling efforts. The COVID-19 pandemic has underscored the critical need for such datasets and modeling techniques to address emerging disease challenges \cite{azzopardi2021call}.

One of the key challenges we encountered was the lack of standardized datasets at the national level, especially in countries like Spain, where management resources are primarily directed toward autonomous regions and municipalities. This implies that, for instance, there are no accurate and uniform datasets describing the number of students and classroom sizes at the national level, as each autonomous region has a different management system with diverse Open Data policies. Although we showed that our methodology was able to address some of these difficulties, we advocate for the development of standardized frameworks, ideally at the international level, but at the very least within each country. Standardization would greatly enhance the usability and comparability of these valuable resources.

It is worth noting that our approach focused on isolated cities, which represents a limitation in capturing the full extent of human mobility and pathogen spreading across large scales. However, this limitation can be addressed by introducing probabilities of case importation into spreading models. Besides, even if our datasets represent static snapshots of reality, their multilayer nature allows for the incorporation of temporal variations, such as daily and weekly work and school schedules, which may impact the weight of contacts in all layers.

In summary, our work offers a foundation for further research and modeling in the field of urban dynamics and infectious disease spread. We hope that these synthetic datasets and our methodologies will contribute to advancing our understanding of complex social systems and their implications for public health and other related domains.

\section{Code availability}
The codes (Jupyter notebooks in Python), used data and generated data are available in a repository, reachable from this link: \url{https://doi.org/10.20350/digitalCSIC/16527}.

\section{Author Contributions \& Competing Interests}
Author contributions: Conceptualization: YM, AA, JR; Software: JR; Data curation: JR; Formal analysis: JR; Supervision: YM, AA; Funding acquisition: YM; Methodology: YM, AA, JR.

Authors declare no competing interests.

\section{Acknowledgements}
The authors acknowledge the Department of University Statistics of the Spanish Ministry of Universities for facilitating the number of registered students at the university per municipality, which is not currently publicly available online. JR, AA, and YM acknowledge support from the Government of Aragon (FONDO–COVID19-UZ-164255). JR is supported by Govern de les Illes Balears through the Vicenç Mut program and the María de Maeztu Excellence Unit 2023-2027 Ref. CEX2021-001164-M, funded by MCIN/AEI/ 10.13039/501100011033. AA acknowledges support through the grant RYC2021-033226-I funded by MCIN/ AEI/ 10.13039/501100011033 and the European Union NextGenerationEU/PRTR. YM was partially supported by the Government of Aragon, Spain, and ERDF "A way of making Europe" through grant E36-20R (FENOL), and by Ministerio de Ciencia e Innovación, Agencia Española de Investigación (MCIN/AEI/ 10.13039/501100011033) Grant No. PID2020-115800GB-I00. The authors acknowledge the use of the computational resources of COSNET Lab at Institute BIFI, funded by Banco Santander through grant Santander-UZ 2020/0274, and by the Government of Aragon (FONDO–COVID19-UZ-164255). JR and AA acknowledge funding from la Caixa Foundation under the project code SR20-00386 (COVID-SHINE). The funders had no role in the study design, data collection, analysis, decision to publish, or preparation of the manuscript.

\bibliographystyle{IEEEtran}
\bibliography{main}
\end{document}